\documentstyle[12pt,fleqn]{article}
\def\ss{\scriptstyle}
\def\sss{\scriptscriptstyle}
\begin{document}

\title{On SL(3,\,{\bf C})-covariant spinor equation\\
and generalized Duffin-Kemmer algebra}

\author{\bf A. V. Solov'yov}

\date{\normalsize\it Division of Theoretical Physics, Faculty of Physics,\\
Moscow State University, Sparrow Hills, 119899, Moscow, Russia}

\maketitle

\begin{abstract}
The SL(3,\,{\bf C})-covariant 9-dimensional equation for a free
3-spinor particle is transformed into the Dirac-like form $(p_{\sss A}
\delta^{\sss A}-M)\Psi=0$. However, the corresponding $\delta$ matrices
do not satisfy the Dirac algebra. It is shown that $\delta^{\sss A}$ lead
to a Finslerian generalization of the Duffin-Kemmer algebra. The Appendix
contains an explicit realization of the $\delta$ matrices.
\end{abstract}

\section*{}

During several years, a theory called {\it binary geometrophysics\/} is being
successfully developed [1]. Within the framework of this theory, in the
article [2], the concept of a {\it 3-spinor\/} as a tensor on the complex
vector space ${\bf C}^3$ with the trilinear antisymmetric ``scalar
multiplication'' ${\bf C}^3\times{\bf C}^3\times{\bf C}^3\to{\bf C}$ was
introduced. Isometries of such a space form a group isomorphic to the group
SL(3,\,{\bf C}) of unimodular complex 3$\times$3 matrices. In the same
article, a homomorphism of SL(3,\,{\bf C}) into the isometry group of a
9-dimensional (over {\bf R}) flat Finslerian space with the metric 
\begin{eqnarray}
\lefteqn{G^{\sss ABC}p_{\sss A}p_{\sss B}p_{\sss C}=(p_0^2 - p_1^2 - p_2^2 -
p_3^2)p_8}\nonumber\\
&&-(p_4^2 + p_5^2 + p_6^2 + p_7^2)p_0 + 2(p_4p_6 + p_5p_7)p_1\nonumber\\
&&+2(p_5p_6 - p_4p_7)p_2 +(p_4^2 + p_5^2 - p_6^2 - p_7^2)p_3
\end{eqnarray}
was constructed. Here ${\ss A},{\ss B},{\ss C}=\overline{0,8}$ and $p_{\sss
A}$ is a real 9-vector. It should be noted that (1) passes into the ordinary
pseudoeuclidean 4-metric $g^{\mu\nu}p_\mu p_\nu=p_0^2-p_1^2-p_2^2-p_3^2$
after the procedure of
the dimensional reduction. From the physical point of view, the expression
(1) is the ``scalar cube'' of a vector in the 9-dimensional momentum space.

Let $i^r$ and $\beta_{\dot s}$ $(r,s=\overline{1,3})$ be 3-spinors of rank
one while $P^{r\dot s}$ be a 3-spinor of rank two such that the matrix
$P\equiv\|P^{r\dot s}\|$ is Hermitian. (As ever, the dotted index means
that the corresponding quantity is transformed according to the complex
conjugate representation of the group SL(3,\,{\bf C}).)\ \ It is not
difficult to show that
\begin{equation}
\det P=G^{\sss ABC}p_{\sss A}p_{\sss B}p_{\sss C}
\end{equation}
when the relations between $P^{r\dot s}$ and $p_{\sss A}$ have the form
\begin{equation}
\left.\begin{array}{@{}r@{\,}c@{\;}l@{\ }r@{\,}c@{\;}l@{\ }r@{\,}c@{\;}l@{}}
P^{1\dot 1}&=&p_0+p_3,&P^{1\dot 2}&=&p_1-ip_2,&P^{1\dot 3}&=&p_4-ip_5\\
P^{2\dot 1}&=&p_1+ip_2,&P^{2\dot 2}&=&p_0-p_3,&P^{2\dot 3}&=&p_6-ip_7\\
P^{3\dot 1}&=&p_4+ip_5,&P^{3\dot 2}&=&p_6+ip_7,&P^{3\dot 3}&=&p_8\end{array}
\right\}\hspace{-2pt}.
\end{equation}

In the work [3], to describe a free 3-spinor particle with a 9-momentum
$p_{\sss A}$, the SL(3,\,{\bf C})-covariant equation
\begin{equation}
\left.\begin{array}{@{}r@{\;}c@{\;}l@{}}
P^{r\dot s}\beta_{\dot s}&=&Mi^r\\
P_{r\dot s}\,i^r&=&M^2\beta_{\dot s}
\end{array}\right\} 
\end{equation}
was proposed, where $P^{r\dot s}$ are defined by (3), $M$ is a real scalar,
and $P_{r\dot s}$ are the cofactors of $P^{r\dot s}$. It is natural to
call $M$ the 9-mass of a particle because substituting the upper equality
of (4) into the lower one (and vice versa) gives a Finslerian analog of
the Klein-Gordon equation for each 3-spinor component: $(G^{\sss
ABC}p_{\sss A}p_{\sss B}p_{\sss C}-M^3)i^r\!\!=\!0$, $(G^{\sss ABC}p_{\sss
A}p_{\sss B}p_{\sss C}-M^3)\beta_{\dot s}\!=\!0$. In [3], it was
also shown that (4) split into the standard 4-dimensional Dirac and
Klein-Gordon equations after the group reduction ${\rm SL}(3,{\bf C})\to{\rm
SL}(2,{\bf C})$.

In general, the equation (4) is {\it quadratic\/} with respect to $p_{\sss
A}$. This follows from (3) and the fact that $P_{r\dot s}$ are
proportional to 2$\times$2 minors of the matrix $P$. The purpose
of the present paper is to transform (4) into the form of an equation
{\it linear\/} with respect to the 9-momentum $p_{\sss A}$. To this end, it
is possible to use the method of Duffin and Kemmer $[4,5]$.

Let us introduce the new variables $\xi_1$, $\xi_2$, \ldots, $\xi_6$ such that
\begin{equation}
\left.\begin{array}{@{}l@{\ \ }l@{}}
P^{2\dot 1}i^1-P^{1\dot 1}i^2=M\xi_1,&P^{2\dot 2}i^1-P^{1\dot 2}i^2=M\xi_4\\
P^{3\dot 1}i^1-P^{1\dot 1}i^3=M\xi_2,&P^{3\dot 2}i^1-P^{1\dot 2}i^3=M\xi_5\\
P^{3\dot 1}i^2-P^{2\dot 1}i^3=M\xi_3,&P^{3\dot 2}i^2-P^{2\dot 2}i^3=M\xi_6
\end{array}\right\}.
\end{equation}
With the help of (5), one can rewrite the lower equality of (4) in the
following form
\begin{equation}
\left.\begin{array}{@{}r@{\;}c@{\;}l@{}}
P^{3\dot 3}\xi_4 - P^{2\dot 3}\xi_5 + P^{1\dot 3}\xi_6&=&M\beta_{\dot 1}\\
-P^{3\dot 3}\xi_1 + P^{2\dot 3}\xi_2 - P^{1\dot 3}\xi_3&=&M\beta_{\dot 2}\\
-P^{3\dot 1}\xi_4 + P^{2\dot 1}\xi_5 - P^{1\dot 1}\xi_6&=&M\beta_{\dot 3}
\end{array}\right\}.
\end{equation}
Thus, (4) is equivalent to the set of equations $P^{r\dot s}\beta_{\dot
s}=Mi^r$, (5)--(6) or, what is the same, to the matrix equation
\begin{equation}
\hat P\Psi=M\Psi,
\end{equation}
where $\Psi=(i^1,i^2,i^3,\beta_{\dot 1},\beta_{\dot 2},\beta_{\dot 3},
\xi_1,\xi_2,\xi_3,\xi_4,\xi_5,\xi_6)^\top$ is the 12-component column and
\begin{equation}
\hat P=\left(\begin{array}{cccc}
\bf 0&P&\bf 0&\bf 0\\
\bf 0&\bf 0&P_1&P_2\\
P_3&\bf 0&\bf 0&\bf 0\\
P_4&\bf 0&\bf 0&\bf 0
\end{array}\right)
\end{equation}
is the 12$\times$12 matrix consisting of the 3$\times$3-blocks:
\begin{equation}
{\bf 0}=\left(
\begin{array}{ccc}
0&0&0\\
0&0&0\\
0&0&0
\end{array}\right),
\end{equation}
\begin{equation}
P=\left(
\begin{array}{ccc}
P^{1\dot 1}&P^{1\dot 2}&P^{1\dot 3}\\
P^{2\dot 1}&P^{2\dot 2}&P^{2\dot 3}\\
P^{3\dot 1}&P^{3\dot 2}&P^{3\dot 3}
\end{array}\right),
\end{equation}
\begin{equation}
P_1=\left(
\begin{array}{ccc}
0&0&0\\
-P^{3\dot 3}&P^{2\dot 3}&-P^{1\dot 3}\\
0&0&0
\end{array}\right),
\end{equation}
\begin{equation}
P_2=\left(
\begin{array}{ccc}
P^{3\dot 3}&-P^{2\dot 3}&P^{1\dot 3}\\
0&0&0\\
-P^{3\dot 1}&P^{2\dot 1}&-P^{1\dot 1}
\end{array}\right),
\end{equation}
\begin{equation}
P_3=\left(
\begin{array}{ccc}
P^{2\dot 1}&-P^{1\dot 1}&0\\
P^{3\dot 1}&0&-P^{1\dot 1}\\
0&P^{3\dot 1}&-P^{2\dot 1}
\end{array}\right),
\end{equation}
\begin{equation}
P_4=\left(
\begin{array}{ccc}
P^{2\dot 2}&-P^{1\dot 2}&0\\
P^{3\dot 2}&0&-P^{1\dot 2}\\
0&P^{3\dot 2}&-P^{2\dot 2}
\end{array}\right).
\end{equation}

Let us raise (8) to the fourth power. The direct calculation shows that
\begin{equation}
{\hat P}^4=(\det P)\hat P. 
\end{equation}
On the other hand, by using (3) and (9)--(14), it is possible to represent
(8) in the form of the linear combination
\begin{equation}
\hat P=p_{\sss A}\delta^{\sss A}
\end{equation}
of nine 12$\times$12 matrices $\delta^{\sss A}$ which are given in the
Appendix. The substitution of (2) and (16) into (15) results in the
identity
\begin{equation}
(p_{\sss A}\delta^{\sss A})^4=G^{\sss ABC}p_{\sss A}p_{\sss B}p_{\sss C}
(p_{\sss D}\delta^{\sss D})
\end{equation}
which is valid for any $p_{\sss A}$; here ${\ss A},{\ss B},{\ss C},{\ss
D}=\overline{0,8}$. It is evident that (17) generalizes the known
4-dimensional identity $(p_\mu\beta^\mu)^3=g^{\mu\nu}p_\mu p_\nu(p_\lambda
\beta^\lambda)$, where $\mu,\nu,\lambda=\overline{0,3}$ and $\beta^\mu$
are the Duffin-Kemmer matrices [4,\,5]. Moreover, it follows from (17)
that the $\delta$ matrices satisfy the conditions
\begin{equation}
\delta^{\sss (A}\delta^{\sss B}\delta^{\sss C}\delta^{\sss D)}=
6\{G^{\sss ABC}\delta^{\sss D}+G^{\sss ABD}\delta^{\sss C}+
G^{\sss ACD}\delta^{\sss B}+G^{\sss BCD}\delta^{\sss A}\},
\end{equation}
where the parentheses denote the symmetrization with respect to all the
superscripts (i.e., the sum over all permutations of ${\ss A}, {\ss B},
{\ss C}, {\ss D}$). In this connection, it is interesting to recall 
important relations of the Duffin-Kemmer algebra
\begin{equation}
\beta^{{\sss(}\mu}\beta^\nu\beta^{\lambda{\sss)}}=2\{g^{\mu\nu}\beta^\lambda +
g^{\lambda\mu}\beta^\nu + g^{\lambda\nu}\beta^\mu\}.
\end{equation} 
It is easy to see the full analogy between the formulae (18) and (19).

Let us return to the equation (7). With the help of~(16), it can be
written finally in the Dirac-like form
\begin{equation}
(p_{\sss A}\delta^{\sss A} - M)\Psi=0,
\end{equation}
where $\delta^{\sss A}$ satisfy the conditions (18). Thus, the purpose of
this paper has been achieved: (4) has been transformed into the $p$-linear
equation (20).

The author is grateful to professor Yu.~S.~Vladimirov for helpful and
encouraging discussions of the obtained results.

\section*{Appendix. The explicit form of the $\boldmath\delta$ matrices}

\[
\delta^{0}=\left(
\begin{array}{cccccccccccc}
 0& 0& 0& 1& 0& 0& 0& 0& 0& 0& 0& 0\\
 0& 0& 0& 0& 1& 0& 0& 0& 0& 0& 0& 0\\
 0& 0& 0& 0& 0& 0& 0& 0& 0& 0& 0& 0\\
 0& 0& 0& 0& 0& 0& 0& 0& 0& 0& 0& 0\\
 0& 0& 0& 0& 0& 0& 0& 0& 0& 0& 0& 0\\
 0& 0& 0& 0& 0& 0& 0& 0& 0& 0& 0& -1\\
 0& -1& 0& 0& 0& 0& 0& 0& 0& 0& 0& 0\\
 0& 0& -1& 0& 0& 0& 0& 0& 0& 0& 0& 0\\
 0& 0& 0& 0& 0& 0& 0& 0& 0& 0& 0& 0\\
 1& 0& 0& 0& 0& 0& 0& 0& 0& 0& 0& 0\\
 0& 0& 0& 0& 0& 0& 0& 0& 0& 0& 0& 0\\
 0& 0& -1& 0& 0& 0& 0& 0& 0& 0& 0& 0
\end{array}
\right),
\]

\[
\delta^{1}=\left(
\begin{array}{cccccccccccc}
 0& 0& 0& 0& 1& 0& 0& 0& 0& 0& 0& 0\\
 0& 0& 0& 1& 0& 0& 0& 0& 0& 0& 0& 0\\
 0& 0& 0& 0& 0& 0& 0& 0& 0& 0& 0& 0\\
 0& 0& 0& 0& 0& 0& 0& 0& 0& 0& 0& 0\\
 0& 0& 0& 0& 0& 0& 0& 0& 0& 0& 0& 0\\
 0& 0& 0& 0& 0& 0& 0& 0& 0& 0& 1& 0\\
 1& 0& 0& 0& 0& 0& 0& 0& 0& 0& 0& 0\\
 0& 0& 0& 0& 0& 0& 0& 0& 0& 0& 0& 0\\
 0& 0& -1& 0& 0& 0& 0& 0& 0& 0& 0& 0\\
 0& -1& 0& 0& 0& 0& 0& 0& 0& 0& 0& 0\\
 0& 0& -1& 0& 0& 0& 0& 0& 0& 0& 0& 0\\
 0& 0& 0& 0& 0& 0& 0& 0& 0& 0& 0& 0
\end{array}
\right),
\]

\[
\delta^{2}=\left(
\begin{array}{cccccccccccc}
 0& 0& 0& 0& -i& 0& 0& 0& 0& 0& 0& 0\\
 0& 0& 0& i& 0& 0& 0& 0& 0& 0& 0& 0\\
 0& 0& 0& 0& 0& 0& 0& 0& 0& 0& 0& 0\\
 0& 0& 0& 0& 0& 0& 0& 0& 0& 0& 0& 0\\
 0& 0& 0& 0& 0& 0& 0& 0& 0& 0& 0& 0\\
 0& 0& 0& 0& 0& 0& 0& 0& 0& 0& i& 0\\
 i& 0& 0& 0& 0& 0& 0& 0& 0& 0& 0& 0\\
 0& 0& 0& 0& 0& 0& 0& 0& 0& 0& 0& 0\\
 0& 0& -i& 0& 0& 0& 0& 0& 0& 0& 0& 0\\
 0& i& 0& 0& 0& 0& 0& 0& 0& 0& 0& 0\\
 0& 0& i& 0& 0& 0& 0& 0& 0& 0& 0& 0\\
 0& 0& 0& 0& 0& 0& 0& 0& 0& 0& 0& 0
\end{array}
\right),
\]

\[
\delta^{3}=\left(
\begin{array}{cccccccccccc}
 0& 0& 0& 1& 0& 0& 0& 0& 0& 0& 0& 0\\
 0& 0& 0& 0& -1& 0& 0& 0& 0& 0& 0& 0\\
 0& 0& 0& 0& 0& 0& 0& 0& 0& 0& 0& 0\\
 0& 0& 0& 0& 0& 0& 0& 0& 0& 0& 0& 0\\
 0& 0& 0& 0& 0& 0& 0& 0& 0& 0& 0& 0\\
 0& 0& 0& 0& 0& 0& 0& 0& 0& 0& 0& -1\\
 0& -1& 0& 0& 0& 0& 0& 0& 0& 0& 0& 0\\
 0& 0& -1& 0& 0& 0& 0& 0& 0& 0& 0& 0\\
 0& 0& 0& 0& 0& 0& 0& 0& 0& 0& 0& 0\\
 -1& 0& 0& 0& 0& 0& 0& 0& 0& 0& 0& 0\\
 0& 0& 0& 0& 0& 0& 0& 0& 0& 0& 0& 0\\
 0& 0& 1& 0& 0& 0& 0& 0& 0& 0& 0& 0
\end{array}
\right),
\]

\[
\delta^{4}=\left(
\begin{array}{@{}cccccccccccc@{}}
 0& 0& 0& 0& 0& 1& 0& 0& 0& 0& 0& 0\\
 0& 0& 0& 0& 0& 0& 0& 0& 0& 0& 0& 0\\
 0& 0& 0& 1& 0& 0& 0& 0& 0& 0& 0& 0\\
 0& 0& 0& 0& 0& 0& 0& 0& 0& 0& 0& 1\\
 0& 0& 0& 0& 0& 0& 0& 0& -1& 0& 0& 0\\
 0& 0& 0& 0& 0& 0& 0& 0& 0& -1& 0& 0\\
 0& 0& 0& 0& 0& 0& 0& 0& 0& 0& 0& 0\\
 1& 0& 0& 0& 0& 0& 0& 0& 0& 0& 0& 0\\
 0& 1& 0& 0& 0& 0& 0& 0& 0& 0& 0& 0\\
 0& 0& 0& 0& 0& 0& 0& 0& 0& 0& 0& 0\\
 0& 0& 0& 0& 0& 0& 0& 0& 0& 0& 0& 0\\
 0& 0& 0& 0& 0& 0& 0& 0& 0& 0& 0& 0
\end{array}
\right),
\]

\[
\delta^{5}=\left(
\begin{array}{cccccccccccc}
 0& 0& 0& 0& 0& -i& 0& 0& 0& 0& 0& 0\\
 0& 0& 0& 0& 0& 0& 0& 0& 0& 0& 0& 0\\
 0& 0& 0& i& 0& 0& 0& 0& 0& 0& 0& 0\\
 0& 0& 0& 0& 0& 0& 0& 0& 0& 0& 0& -i\\
 0& 0& 0& 0& 0& 0& 0& 0& i& 0& 0& 0\\
 0& 0& 0& 0& 0& 0& 0& 0& 0& -i& 0& 0\\
 0& 0& 0& 0& 0& 0& 0& 0& 0& 0& 0& 0\\
 i& 0& 0& 0& 0& 0& 0& 0& 0& 0& 0& 0\\
 0& i& 0& 0& 0& 0& 0& 0& 0& 0& 0& 0\\
 0& 0& 0& 0& 0& 0& 0& 0& 0& 0& 0& 0\\
 0& 0& 0& 0& 0& 0& 0& 0& 0& 0& 0& 0\\
 0& 0& 0& 0& 0& 0& 0& 0& 0& 0& 0& 0
\end{array}
\right),
\]

\[
\delta^{6}=\left(
\begin{array}{cccccccccccc}
 0& 0& 0& 0& 0& 0& 0& 0& 0& 0& 0& 0\\
 0& 0& 0& 0& 0& 1& 0& 0& 0& 0& 0& 0\\
 0& 0& 0& 0& 1& 0& 0& 0& 0& 0& 0& 0\\
 0& 0& 0& 0& 0& 0& 0& 0& 0& 0& -1& 0\\
 0& 0& 0& 0& 0& 0& 0& 1& 0& 0& 0& 0\\
 0& 0& 0& 0& 0& 0& 0& 0& 0& 0& 0& 0\\
 0& 0& 0& 0& 0& 0& 0& 0& 0& 0& 0& 0\\
 0& 0& 0& 0& 0& 0& 0& 0& 0& 0& 0& 0\\
 0& 0& 0& 0& 0& 0& 0& 0& 0& 0& 0& 0\\
 0& 0& 0& 0& 0& 0& 0& 0& 0& 0& 0& 0\\
 1& 0& 0& 0& 0& 0& 0& 0& 0& 0& 0& 0\\
 0& 1& 0& 0& 0& 0& 0& 0& 0& 0& 0& 0
\end{array}
\right),
\]

\[
\delta^{7}=\left(
\begin{array}{cccccccccccc}
 0& 0& 0& 0& 0& 0& 0& 0& 0& 0& 0& 0\\
 0& 0& 0& 0& 0& -i& 0& 0& 0& 0& 0& 0\\
 0& 0& 0& 0& i& 0& 0& 0& 0& 0& 0& 0\\
 0& 0& 0& 0& 0& 0& 0& 0& 0& 0& i& 0\\
 0& 0& 0& 0& 0& 0& 0& -i& 0& 0& 0& 0\\
 0& 0& 0& 0& 0& 0& 0& 0& 0& 0& 0& 0\\
 0& 0& 0& 0& 0& 0& 0& 0& 0& 0& 0& 0\\
 0& 0& 0& 0& 0& 0& 0& 0& 0& 0& 0& 0\\
 0& 0& 0& 0& 0& 0& 0& 0& 0& 0& 0& 0\\
 0& 0& 0& 0& 0& 0& 0& 0& 0& 0& 0& 0\\
 i& 0& 0& 0& 0& 0& 0& 0& 0& 0& 0& 0\\
 0& i& 0& 0& 0& 0& 0& 0& 0& 0& 0& 0
\end{array}
\right),
\]

\[
\delta^{8}=\left(
\begin{array}{cccccccccccc}
 0& 0& 0& 0& 0& 0& 0& 0& 0& 0& 0& 0\\
 0& 0& 0& 0& 0& 0& 0& 0& 0& 0& 0& 0\\
 0& 0& 0& 0& 0& 1& 0& 0& 0& 0& 0& 0\\
 0& 0& 0& 0& 0& 0& 0& 0& 0& 1& 0& 0\\
 0& 0& 0& 0& 0& 0& -1& 0& 0& 0& 0& 0\\
 0& 0& 0& 0& 0& 0& 0& 0& 0& 0& 0& 0\\
 0& 0& 0& 0& 0& 0& 0& 0& 0& 0& 0& 0\\
 0& 0& 0& 0& 0& 0& 0& 0& 0& 0& 0& 0\\
 0& 0& 0& 0& 0& 0& 0& 0& 0& 0& 0& 0\\
 0& 0& 0& 0& 0& 0& 0& 0& 0& 0& 0& 0\\
 0& 0& 0& 0& 0& 0& 0& 0& 0& 0& 0& 0\\
 0& 0& 0& 0& 0& 0& 0& 0& 0& 0& 0& 0
\end{array}
\right).
\]

\end{document}